\documentclass[aps,prd,onecolumn,groupedaddress,showpacs,nofootinbib,amssymb,balancelastpage]{revtex4}
\usepackage{graphicx,bm,xcolor}
\usepackage{amsmath}
\usepackage{amssymb}
\usepackage{amsfonts}
\usepackage{cases}
\usepackage{cancel}

\newcommand\be{\begin{equation}}
\newcommand\ee{\end{equation}}
\newcommand\bea{\begin{eqnarray}}
\newcommand\eea{\end{eqnarray}}

\newcommand\e{\mathrm{e}}

\begin{document}

\title{Relativistic Stars in dRGT Massive Gravity}
\author{Taishi Katsuragawa$^1$, Shin'ichi Nojiri$^{1,2}$, Sergei D.
Odintsov$^{3,4,5}$, Masashi Yamazaki$^1$ }
\affiliation{
$^{1)}$ Department of Physics, Nagoya University, Nagoya 464-8602, Japan \\
$^{2)}$ Kobayashi-Maskawa Institute for the Origin of Particles and the Universe, Nagoya University, Nagoya 464-8602, Japan \\
$^{3)}$ Institut de Ciencies de lEspai (IEEC-CSIC), Campus UAB, Carrer de Can Magrans, s/n, 08193 Cerdanyola del Valles, Barcelona, Spain\\
$^{4)}$ ICREA, Passeig LluAs Companys, 23, 08010 Barcelona, Spain \\
$^{5)}$ Tomsk State Pedagogical University, 634061 Tomsk, Russia}

\begin{abstract}
We study relativistic stars in the simplest model of the de
Rham-Gabadadze-Tolley massive gravity
which describes the massive graviton without ghost propagating mode.
We consider the hydrostatic equilibrium,
and obtain the modified Tolman-Oppenheimer-Volkoff equation and the
constraint equation
coming from the potential terms in the gravitational action.
We give analytical and numerical results for quark and neutron stars and
discuss the deviations compared with General Relativity and $F(R)$ gravity.
It is shown that the theory under investigation leads to small deviation from
the General Relativity
in terms of density profiles and mass-radius relation.
Nevertheless, such deviation may be observable in future astrophysical
probes.
\end{abstract}


\maketitle

\section{Introduction}

Late-time accelerated expansion of the Universe has been confirmed by
several independent observations
\cite{Perlmutter:1998np,Riess:1998cb,Spergel:2003cb,Spergel:2006hy,Komatsu:2008hk,Komatsu:2010fb,Tegmark:2003ud,Seljak:2004xh,Eisenstein:2005su}.
In order to explain the accelerated expansion,
we need to include new energy sources,
which are known as the Dark Energy (DE).
The simplest example of the DE is the cosmological constant $\Lambda$.
Furthermore, by including the Cold Dark Matter (CDM) besides the Standard
Model (SM) particles,
we obtain the well-known $\Lambda$CDM model, which successfully describes
the current epoch of the Universe.
The $\Lambda$CDM model is, however, merely one of the phenomenological
models,
and from the theoretical point of view, it suffers from several
theoretical problems.
For example, we need to theoretically explain the very large disagreement
between the theoretically estimated values of cosmological constant and
the observational value,
and the ratio of ordinary matters and CDM with respect to DE in the
current epoch.

Since one may regard that the inclusion of the cosmological constant could
be a minimal extension of the General Relativity,
we can also consider the other extensions of the General Relativity
and investigate the possibility that these extended theories could
describe the real universe.
As non-minimal extensions of the General Relativity to explain the current
expansion without cosmological constant,
the modified gravitational theories have been proposed and investigated well
(for review, see, for instance,
\cite{Nojiri:2006ri,Faraoni,Bamba:2012cp,Khoury,Capozziello:2011et}).
In order to establish such a new gravitational theory,
it is important to study the cosmological models and compare them with the observational data.
We should note that the modified gravitational theories could be also constrained by the astrophysical observations,
for example, those of the compact stars.

Recently, massive and compact neutron stars
whose mass is $M_\mathrm{NS} \sim 2M_{\odot}$ ($M_{\odot}$ is the solar mass) were found
\cite{Barziv:2001ad,Rawls:2011jw,Nice:2005fi,Demorest:2010bx}.
It could be hardly understood in the framework of the General Relativity
and hadron physics so far
if one uses the stellar matter equations of state which are comfortable in
astrophysics and hadron physics.
Thus, there could be two points of view to explain the massive neutron stars:
One is from particle physics side which requires not-well justified
phenomenological change of equation of state and
another is the gravitational physics side while using the convenient equation of state for stellar matter.
The size of the compact objects is determined by the balance between
degeneracy force and gravitational force.
In order to explain the massive neutron star,
three approaches seem to be reasonable:
(i) the repulsive force is stronger than that is realized with the
standard equations of state,
(ii) the attractive force is weaker than that is predicted in the General
Relativity,
(iii) we accept both cases (i) and (ii) simultaneously.
From the point of view of (i) based on the hadron physics,
it was suggested that equations of state could be modified by introducing
the new interactions
\cite{Miyatsu:2013yta,Tsubakihara:2012ic}.
From the point of view of (ii) based on the gravitational physics,
it has been suggested that some models of $F(R)$ gravity can explain the
massive and compact neutron stars
\cite{Capo1,Capozziello:2015yza,Kostas}.
In this work, we stand on the viewpoint of case (ii),
and study if the compact objects, quark stars and neutron stars, are realized
and how the internal structure of compact objects is deviated from that in
the General Relativity
if we assume the massive gravity coupled with the matter which is
described by the standard equations of state.

The de Rham-Gabadadze-Tolley (dRGT) massive gravity
\cite{deRham:2010ik,deRham:2010kj,Hassan:2011hr} (for review, see
\cite{deRham})
is the ghost-free theory with interacting massive spin-2 field.
The basic idea of the massive spin-$2$ field theory was proposed by Fierz
and Pauli in Ref.~\cite{Fierz:1939ix},
where the consistent free massive spin-$2$ theory was given
by adding a tuned mass term to free massless spin-$2$ field theory on flat
space-time was considered.
It was, however, shown that the Fierz-Pauli theory cannot recover the
General Relativity in the massless limit,
due to so-called the van Dam-Veltman-Zakharov (vDVZ) discontinuity
\cite{vanDam:1970vg,Zakharov:1970cc}.
After the discovery of the vDVZ discontinuity, in order to avoid this
problem,
the Fierz-Pauli theory was extended to an interacting theory by replacing
the kinetic terms with the Ricci scalar.
As a result, the vDVZ discontinuity can be screened by a non-linear effect
which is called the Vainshtein mechanism \cite{Vainshtein:1972sx}.
Nevertheless, the non-linear terms generate a ghost called the
Boulware-Deser (BD) ghost \cite{Boulware:1974sr}.
The problem of the BD ghost mode had been discussed for a long time,
and the problem has been solved finally as the dRGT massive gravity by
introducing a new form of mass terms.

The dRGT massive gravity is considered to be able to avoid the constraint from
the Solar System and terrestrial experiments thanks to the Vainshtein mechanism,
where the non-linear effects hide the extra degree of freedom coupled with
the matter source.
At the same time, the massive graviton leads to the modification of
long-range gravitational force
because the gravitational potential is modified to be the Yukawa-type
potential,
where the scale of modification is characterized by that of graviton mass.
Thus, one may expect that the mass of massive graviton could be comparable
to the cosmological constant,
which could explain the accelerated expansion of the Universe without
introducing the cosmological constant
\cite{D'Amico:2011jj,Gumrukcuoglu:2011ew,Comelli:2011zm,Langlois:2012hk,Kobayashi:2012fz}.

As we regard the dRGT massive gravity as an alternative theory of gravity,
it is interesting to apply this theory to astrophysical phenomena as well
as to the accelerated expansion of the Universe.
It is very difficult to construct the general framework which quantifies
the deviations
from the predictions of the General Relativity in strong-gravity field
because the non-perturbative effects depend on the detail of each theories
and
parametric treatment is not suitable
\cite{Psaltis:2008bb}.
Furthermore, it it significant if we could conclude
that astrophysical and cosmological applications are compatible with
observations in specific theory of modified gravity.
From the above reasons, it is indispensable to study the compact objects in the
dRGT massive gravity in the same way as it was done in $F(R)$ gravity,
as astrophysical test of the massive gravity in strong-gravity regime.

This paper is organized as follows:
In Section II, we give a brief review of the dRGT massive gravity and
derive the equations of motion.
In Section III, we consider the spherically-symmetric stellar metric and
derive the TOV equations in minimal model of the dRGT massive gravity.
It is shown that one constraint equation coming from the potential terms
in the gravitational action appears.
It leads to the explicit difference from the case of the General Relativity.
In section IV, we present the numerical analysis for the quark star and
neutron star
for some convenient equations of state.
In particular, mass-central density and mass-radius relation are
numerically analyzed.
In section V, we summarize the obtained results for relativistic stars and
discuss the difference of the massive gravity from the General Relativity
and the $F(R)$ gravity.

\section{The Action and Equation of Motion in dRGT Massive Gravity}

In this section, we give a brief review of the dRGT massive gravity and
derive the equation of motion.
The action of the dRGT massive gravity \cite{Hassan:2011hr} is given by
\begin{align}
S_\mathrm{dRGT} =&
\frac{1}{2\kappa^{2}} \int d^{4}x \sqrt{-\mathrm{det}(g)} \left[
R - 2m^{2}_{0} \sum^{4}_{n=0} \beta_{n}e_{n} \left( \sqrt{g^{-1}f} \right)
\right]
+ S_{\mathrm{matter}}\, .
\label{the action}
\end{align}
Here, $g_{\mu \nu}$ and $f_{\mu \nu}$ are dynamical and reference metrics,
respectively,
and $\kappa$ is the gravitational coupling given in terms of the Newton
constant of gravitation $G$,
$\kappa^{2}=8\pi G$.
In (\ref{the action}), the coefficients $\beta_{n}$s and $m_{0}$ are free
parameters.
The matrix $\sqrt{g^{-1}f}$ is defined as the square root of $g^{\mu
\rho}f_{\rho \nu}$, that is,
\begin{align}
\left( \sqrt{g^{-1}f} \right)^{\mu}_{\ \rho} \left( \sqrt{g^{-1}f}
\right)^{\rho}_{\ \nu} = g^{\mu \rho}f_{\rho \nu}\, .
\label{sqrtfg}
\end{align}
For general matrix $\mathbf{X}$, $e_{n}(\mathbf{X})$s are defined as
polynomials of the eigenvalues of $X$:
\begin{align}
e_{0}(\mathbf{X}) =& 1\, , \quad
e_{1}(\mathbf{X}) = [\mathbf{X}] \, , \nonumber \\
e_{2}(\mathbf{X}) =& \frac{1}{2} \left( [\mathbf{X}]^{2} - [\mathbf{X}^{2}]
\right) \, , \nonumber \\
e_{3}(\mathbf{X}) =& \frac{1}{6}
\left( [\mathbf{X}]^{3} - 3[\mathbf{X}][\mathbf{X}^{2}] + 2[\mathbf{X}^{3}]
\right) \, , \nonumber \\
e_{4}(\mathbf{X}) =& \frac{1}{24}
\left( [\mathbf{X}]^{4} - 6[\mathbf{X}]^{2}[\mathbf{X}^{2}] +
3[\mathbf{X}^{2}]^{2}
+8[\mathbf{X}][\mathbf{X}^{3}] - 6[\mathbf{X}^{4}] \right)
= \mathrm{det}(\mathbf{X}) \, , \nonumber \\
e_{k}(\mathbf{X}) =& 0 \quad \mbox{for} \ \ k>4 \, , \label{e_n}
\end{align}
where the square brackets denote traces of the matrices, that is,
$[X]=X^{\mu}_{\ \mu}$.
For conventional notations in this paper, hereafter, we denote the
determinant of a matrix $A$ as $\mathrm{det}(A)$,
and $\sqrt{A}$ represents a matrix which is the square root of $A$.

We should note that non-dynamical tensor is required in order to describe
the massive spin-$2$ field
because we cannot construct the potential terms without derivatives only
by using $g_{\mu \nu}$.
We may consider the invariants which consist of $g_{\mu \nu}$, for example,
$g^{2}_{\mu \nu}$ or $g^{\mu}_{\ \mu}$,
but they are constants and corresponding to the cosmological constant.
We should also note that $e_{4} ( \sqrt{g^{-1}f} )$ can be ignored when we
study the dynamics because
\begin{align}
\sqrt{-\mathrm{det}(g)} \, e_{4} ( \sqrt{g^{-1}f} ) = \sqrt{-\mathrm{det}(f)}
\, ,
\end{align}
which is non-dynamical since $f_{\mu \nu}$ is non-dynamical tensor, and
does not appear in the equation of motion.
By the variation of $g_{\mu \nu}$ in Eq.~(\ref{the action}),
we obtain the following equation of motion:
\begin{align}
0 =& R_{\mu \nu}(g) - \frac{1}{2}R(g)g_{\mu \nu} \nonumber \\
&+ \frac{1}{2} m^{2}_{0}
\sum^{3}_{n=0}(-1)^{n}\beta_{n}
 \left[ g_{\mu \lambda}Y^{\lambda}_{(n) \nu}(\sqrt{g^{-1}f})
+ g_{\nu \lambda}Y^{\lambda}_{(n) \mu}(\sqrt{g^{-1}f}) \right]
- \kappa^{2}T_{\mu \nu} \, .
\label{geq}
\end{align}
Here, for a matrix $\mathbf{X}$, $Y_{n}(\mathbf{X})$s are defined by
\begin{align}
Y^{\lambda}_{(n) \nu}(\mathbf{X})=\sum^{n}_{r=0} (-1)^{r} \left( X^{n-r}
\right)^{\lambda}_{\ \nu} e_{r}(\mathbf{X})\, ,
\end{align}
or explicitly,
\begin{align}
Y_{0}(\mathbf{X}) =& \mathbf{1}\, , \quad
Y_{1}(\mathbf{X}) = \mathbf{X}- \mathbf{1}[\mathbf{X}]\, , \quad
Y_{2}(\mathbf{X}) = \mathbf{X}^{2} - \mathbf{X}[\mathbf{X}]
+ \frac{1}{2} \mathbf{1} \left( [\mathbf{X}]^{2} - [\mathbf{X}^{2}]
\right)\, ,
\nonumber \\
Y_{3}(\mathbf{X}) =& \mathbf{X}^{3} - \mathbf{X}^{2}[\mathbf{X}]
+ \frac{1}{2} \mathbf{X} \left( [\mathbf{X}]^{2} - [\mathbf{X}^{2}] \right)
- \frac{1}{6} \mathbf{1} \left( [\mathbf{X}]^{3} -
3[\mathbf{X}][\mathbf{X}^{2}] + 2[\mathbf{X}^{3}] \right)\, .
\end{align}
Note that since $e_{n}$s are written in terms of the trace of $g^{-1}f$,
the following formula about the variation of the trace could be useful,
\begin{align}
\delta \mathrm{tr}\left( (\sqrt{g^{-1}f} )^{n} \right)
= \frac{n}{2} \mathrm{tr} \left( g (\sqrt{g^{-1}}f)^{n} \delta g^{-1}
\right)\, .
\end{align}
Then, we obtain
\begin{align}
\frac{2}{\sqrt{- \mathrm{det}(g)}}
\delta_{g} \left( \sqrt{- \mathrm{det}(g)}e_{n}(\sqrt{g^{-1}f}) \right)
=\sum^{n}_{r=0}(-1)^{r+1} \mathrm{tr} \left( g (\sqrt{g^{-1}f})^{r} \delta
g^{-1} \right) e_{n-r}(\sqrt{g^{-1}f})\, ,
\end{align}
and the third term in Eq.~(\ref{geq}) is symmetrized with respect to the
indices $\mu$ and $\nu$.
If the metrics $g$ and $f$ are diagonal,
the matrix $\sqrt{g^{-1}f}$ is symmetric and the equation of motion is
written as
\begin{align}
G_{\mu \nu} + m^{2}_{0}I_{\mu \nu} = \kappa^{2}T_{\mu \nu}\, ,
\label{eom}
\end{align}
where $G_{\mu \nu}$ is the Einstein tensor, and we define the sum of
interaction terms $I_{\mu \nu}$ as follows:
\begin{align}
I_{\mu \nu}= \sum^{3}_{n=0}(-1)^{n}\beta_{n} g_{\mu
\lambda}Y^{\lambda}_{(n) \nu}(\sqrt{g^{-1}f}) \, .
\end{align}
In Eq.~(\ref{eom}), $T_{\mu \nu}$ is the energy-momentum tensor,
and we assume that matter is minimally coupled to gravity
in order to avoid the ghost problem due to non-minimal matter couplings
\cite{deRham:2014fha,deRham:2014naa}.

\section{Modified TOV equations}

\subsection{Ansatz}

In this section, we study the static and spherical equations of motion
with the perfect fluid in hydrostatic equilibrium.
It is called the TOV equation in the General Relativity.
At first, we calculate the curvature and the interaction terms for
spherically symmetric case,
and check how the TOV equation is modified in the dRGT massive gravity.

For the dynamical metric $g_{\mu \nu}$ and reference metric $f_{\mu \nu}$,
we assume the static and spherically symmetric ansatz in polar coordinate
system,
\begin{align}
g_{\mu \nu}dx^{\mu}dx^{\nu} =&
- \e^{2\phi}dt^{2} + \e^{2\lambda} d\rho^{2} + D^{2}(\rho) \left(
d\theta^{2} + \sin^{2}\theta d \varphi^{2} \right) \, , \\
f_{\mu \nu}dx^{\mu}dx^{\nu} =&
- h(\rho)dt^{2} + h^{-1}(\rho)d\rho^{2} + \rho^{2} \left( d\theta^{2} +
\sin^{2}\theta d \varphi^{2} \right) \, ,
\end{align}
where $\rho$ is radial coordinate, and $\phi$ and $\lambda$ are functions of $\rho$, 
$\phi = \phi(\rho)$ and $\lambda = \lambda(\rho)$.
We note that 
we do not consider the general class of the reference metric 
but specific one which is inspired by static and spherically symmetric solution in the General Relativity.
$h(\rho)$ is a function of $\rho$, for example, $h(r) = 1 -\frac{2M}{r}$ for the Schwarzschild-type metric.
We also assume that the center of the space-time described by $f_{\mu \nu}$ locates at 
the center of physical space-time described by $g_{\mu \nu}$ for simplicity.

In order to compare the difference between the General Relativity and the dRGT massive gravity,
we change the form of above ansatz as follows:
We define the new variable $r$ so that $D(\rho) = r^{2}$, 
which can be solved with respect to $\rho$, $\rho = \chi(r)$, 
and we find
\begin{align}
g_{\mu \nu}dx^{\mu}dx^{\nu} =&
- \e^{2\phi}dt^{2} + \e^{2\lambda} dr^{2} + r^{2}\left( d\theta^{2} +
\sin^{2}\theta d \varphi^{2} \right) \, , \\
f_{\mu \nu}dx^{\mu}dx^{\nu} =&
- h(r)dt^{2} + h^{-1}(r) \left( \chi^{\prime}(r) \right)^{2} dr^{2} +
\chi^{2}(r) \left( d\theta^{2} + \sin^{2}\theta d \varphi^{2} \right) \, .
\end{align}
We note that the scalar function $\chi(r)$ is corresponding to the degree
of freedom of the Stukelberg field.
The general coordinate transformation invariance is broken in the massive
gravity, but it can be restored by changing the Stukelberg field.
In our case, the radial coordinate is chosen
so that the dynamical metric identified with physical space-time is
treated in same procedure as the TOV equation.

For the above ansatz,
we obtain the non-vanishing components of the Ricci tensor and the Ricci
scalar as follows:
\begin{align}
&R_{tt} = \left[ \phi^{\prime \prime} + (\phi^{\prime})^{2} -
\phi^{\prime}\lambda^{\prime}
+ \frac{2\phi^{\prime}}{r} \right]\e^{2(\phi-\lambda)} \, , \\
&R_{rr} = -\phi^{\prime \prime} - (\phi^{\prime})^{2} +
\phi^{\prime}\lambda^{\prime}
+\frac{2\lambda^{\prime}}{r} \, ,\\
&R_{\theta \theta} = - ( 1 - \lambda^{\prime}r +
\phi^{\prime}r)\e^{-2\lambda}
+ 1 \, , \\
&R_{\varphi \varphi} = \sin^{2}\theta R_{\theta \theta} \, ,\\
&R=2 \left[ -\phi^{\prime \prime} - (\phi^{\prime})^{2} +
\phi^{\prime}\lambda^{\prime}
+ \frac{2\lambda^{\prime}}{r} - \frac{2\phi^{\prime}}{r} - \frac{1}{r^{2}}
\right]\e^{-2\lambda}
+ \frac{2}{r^{2}} \, ,
\end{align}
and the non-vanishing components of the Einstein tensor as follows:
\begin{align}
G_{tt} =&
\frac{1}{r^{2}}\e^{2\phi} -
\frac{1-2\lambda^{\prime}r}{r^{2}}\e^{2\phi-2\lambda} \, , \\
G_{rr}=& -\frac{1}{r^{2}}\e^{2\lambda} + \frac{1+2 \phi^{\prime}r}{r^{2}} \,
,\\
G_{\theta\theta}=&
r^{2} \left[ \phi^{\prime \prime} + (\phi^{\prime})^{2} -
\phi^{\prime}\lambda^{\prime}
- \frac{\lambda^{\prime}}{r} + \frac{\phi^{\prime}}{r}
\right]\e^{-2\lambda} \,
, \\
G_{\phi\phi}=& r^{2}\sin^{2}\theta
\left[ \phi^{\prime \prime} + (\phi^{\prime})^{2} -
\phi^{\prime}\lambda^{\prime}
- \frac{\lambda^{\prime}}{r} + \frac{\phi^{\prime}}{r}
\right]\e^{-2\lambda} \, .
\end{align}

\subsection{The minimal model with flat reference metric}
Next, we calculate the interaction terms in Eq.~(\ref{eom}) to obtain the
modified TOV equation in massive gravity.
It is, however, not so easy to study the all cases with different
parameters and different reference metrics
because it is impossible to obtain the general solution for all models in
the dRGT massive gravity.
In this subsection, thus, we specify the parameters $\beta_{n}$ and
reference metric
and  study the modified TOV equation.

First, we introduce a minimal model of the dRGT massive gravity
where the parameters $\beta_{n}$ are chosen as follows:
\begin{align}
\beta_{0} = 3\, , \quad \beta_{1} = -1\, , \quad \beta_{2} = 0\, , \quad
\beta_{3} = 0 \, .
\label{minimal}
\end{align}
Here, we should note that
the interaction terms in (\ref{the action}) can be expressed in terms of
another variable $K=\sqrt{g^{-1}f} - 1$,
\begin{align}
\sum^{3}_{n=0} \beta_{n}e_{n} \left( \sqrt{g^{-1}f} \right) = \sum^{3}_{n=0}
\alpha_{n}e_{n} \left(K \right) \, .
\end{align}
The parameters $\alpha_{n}$ are related to $\beta_{n}$ by the following
relation,
\begin{align}
\beta_{i} = (4-i)! \sum^{4}_{n=i} \frac{(-1)^{n+i}}{(4-n)!(n-i)!}
\alpha_{n} \, .
\end{align}
Then, if we require the flat solution and the recovery of the covariant
Fierz-Pauli action in the limit where the gravitational coupling vanishes,
the general action of massive gravity with parameters $\beta_{n}$ are
reduced to be a 2-parameter family
with two parameters $\alpha_{3}$ and $\alpha_{4}$,
where the minimal model corresponds to $(\alpha_{3}, \alpha_{4}) = (1,1)$.

So, we find that the interaction terms in minimal model are given as follows:
\begin{align}
I_{tt}
&=g_{tt}\left( \beta_{0}Y^{t}_{(0) t} -\beta_{1}Y^{t}_{(1) t} 
+ \beta_{2}Y^{t}_{(2) t} -\beta_{3}Y^{t}_{(3) t} \right) \nonumber \\
&= - \e^{2 \phi (r)} \left( 3- \frac{2 \chi (r)}{r} - \frac{ \chi^{\prime} (r)}{\sqrt{h(r)}} \e^{-\lambda (r)} \right)
\label{int1} \, ,
\end{align}
\begin{align}
I_{rr}
&=g_{rr}\left( \beta_{0}Y^{r}_{(0) r} -\beta_{1}Y^{r}_{(1) r} 
+ \beta_{2}Y^{r}_{(2) r} -\beta_{3}Y^{r}_{(3) r} \right) \nonumber \\
&=\e^{2 \lambda (r)} \left( 3 -\frac{2 \chi (r)}{r} - \sqrt{h(r)} \e^{-\phi (r)} \right)
\label{int2} \, ,
\end{align}
\begin{align}
I_{\theta \theta}
&=g_{\theta \theta}\left( \beta_{0}Y^{\theta}_{(0) \theta}
-\beta_{1}Y^{\theta}_{(1) \theta}
+\beta_{2}Y^{\theta}_{(2) \theta} -\beta_{3}Y^{\theta}_{(3) \theta}
\right) \nonumber \\
&=r^{2} \left( 3 - \frac{\chi (r)}{r} 
- \frac{\chi^{\prime}(r)}{\sqrt{h(r)}} \e^{-\lambda (r)} - \sqrt{h(r)} \e^{-\phi (r)} \right)
\label{int3} \, ,
\end{align}
\begin{align}
I_{\phi \phi}
&=g_{\phi \phi}\left( \beta_{0}Y^{\phi}_{(0) \phi} -\beta_{1}Y^{\phi}_{(1) \phi}
+ \beta_{2}Y^{\phi}_{(2) \phi} - \beta_{3}Y^{\phi}_{(3) \phi} \right)
\nonumber \\
&=r^{2} \sin^2 \theta \left( 3 - \frac{\chi (r)}{r} 
- \frac{\chi^{\prime} (r)}{\sqrt{h(r)}} \e^{-\lambda (r)} - \sqrt{h(r)} \e^{-\phi (r)} \right)
\label{int4} \, .
\end{align}
Now, we derive the equation of motion (\ref{eom}) in the minimal model.
For matter field, we consider a perfect fluid with the following
energy-momentum tensor
\begin{align}
&T_{\mu \nu} = \mathrm{diag}\left(\e^{2\phi} \rho, \e^{2\lambda}P, r^{2}P, r^{2}\sin^{2}\theta P\right)\, .
\end{align}
Then, one obtains $(t,t)$, $(r,r)$, and $(\theta, \theta), (\varphi,
\varphi)$ components as follows:
\begin{align}
- 8 \pi  G \rho (r) =
- \frac{1}{r^2}
+ \frac{1 - 2 \lambda^{\prime}(r)r}{r^{2}} \e^{-2 \lambda (r)}
+ m^{2}_{0}
\left( 3 - \frac{2 \chi (r)}{r}  - \frac{\chi^{\prime} (r)}{\sqrt{h(r)}} \e^{-\lambda (r)}  \right) \, , 
\end{align}
\begin{align}
8 \pi G P(r) =
- \frac{1}{r^2} 
+ \frac{1 + 2 \phi^{\prime}(r) r}{r^2} \e^{-2 \lambda (r)}
+ m^{2}_{0} \left( 3 - \frac{2 \chi (r)}{r} - \sqrt{h(r)} \e^{-\phi (r)} \right) \, , 
\end{align}
\begin{align}
8 \pi G P(r) =&
\e^{-2 \lambda (r)} \left( \phi^{\prime \prime} +  (\phi^{\prime})^{2} - \phi^{\prime}\lambda^{\prime}
- \frac{\lambda^{\prime}}{r} + \frac{\phi^{\prime}}{r} \right)
+ m^{2}_{0} \left( 3 - \frac{\chi (r)}{r}  - \frac{ \chi^{\prime} (r)}{\sqrt{h(r)}} \e^{-\lambda (r)} 
- \sqrt{h(r)} \e^{-\phi (r)}  \right) \, .
\end{align}
We should note that $(\theta, \theta)$ or $(\varphi, \varphi)$ component
of field equation plays a crucial role in contrast to the General
Relativity.
In the static and spherically symmetric case, the Einstein equation leads
to two non-trivial equations,  $(t,t)$ and $(r,r)$ components.
However, in the massive gravity, we need to take the $(\theta, \theta)$ or
$(\varphi, \varphi)$ into account
because the degrees of freedom increase by introducing the second metric
$f_{\mu \nu}$.

Next, we fix the reference metric $f_{\mu \nu}$.
For simplicity, we assume that $h(r)=1$ in the reference metric $f_{\mu
\nu}$,
that is, we consider the Minkowski metric as reference one with an extra
arbitrary function $\chi(r)$.
\begin{align}
f_{\mu \nu}  dx^{\mu} dx^{\nu}
= - dt^{2} + \left( \chi^{\prime}(r) \right)^{2} dr^{2} + \chi^{2}(r)
\left(  d\theta^{2} + \sin^{2}d \varphi^{2} \right) \, ,
\quad h(r)=1 \, .
\end{align}
Note that the choice of the reference metric requires special attention
for cosmological applications.
If one considers the simple FRW ansatz for $g_{\mu \nu}$ with the
Minkowski metric for $f_{\mu \nu}$,
one cannot obtain non-trivial flat FRW cosmology
\cite{D'Amico:2011jj,Gumrukcuoglu:2011ew}.
However, in this case we choose the flat reference metric
because it is better to limit the number of free parameters for the
numerical calculation later.
As a result, in our model, the free parameter is only the graviton mass.
Furthermore, we fix the graviton mass by choosing the mass to be the
cosmological scale.

In this case, the equation of motion are
$(t,t)$, $(r,r)$, and $(\theta, \theta), (\varphi, \varphi)$ components of
equation of motion are given by
\begin{align}
- 8 \pi  G \rho (r) =
- \frac{1}{r^2}
+  \frac{1-2\lambda^{\prime}(r)r}{r^{2}} \e^{-2 \lambda (r)}
+ m^{2}_{0}
\left( 3 - \frac{2 \chi (r)}{r} - \chi^{\prime}(r) \e^{-\lambda (r)}  \right) \, .
\label{tov1-1}
\end{align}
\begin{align}
8 \pi G P(r) =
- \frac{1}{r^2}
+ \frac{1 + 2 \phi^{\prime}(r) r}{r^2} \e^{-2 \lambda (r)}
+ m^{2}_{0} \left( 3 - \frac{2 \chi (r)}{r} - \e^{-\phi (r)}  \right) \, , 
\label{tov2-1}
\end{align}
\begin{align}
8 \pi G P(r) =&
\e^{-2 \lambda (r)} \left(
\phi^{\prime \prime} +  (\phi^{\prime})^{2} - \phi^{\prime}\lambda^{\prime}
- \frac{\lambda^{\prime}}{r} + \frac{\phi^{\prime}}{r} \right)
+ m^{2}_{0}
\left( 3 - \frac{\chi (r)}{r}  - \chi^{\prime} (r) \e^{-\lambda (r)} - \e^{-\phi (r)} \right) \, .
\label{tov3-1}
\end{align}

Finally, we change the variables and rewrite the field equations
Eqs.~(\ref{tov1-1}) and (\ref{tov2-1}).
Let us define the a variable $M(r)$, which is called mass parameter, as follows:
\begin{align}
\e^{-2\lambda(r)} = 1 - \frac{2GM(r)}{r} \, ,
\label{Schw}
\end{align}
because we expect that external space-time is described by the
asymptotically Schwarzschild metric.
Differentiating the above relation (\ref{Schw}) with respect to $r$, we
obtain the following equation
\begin{align}
-\frac{2G}{r^{2}}M^{\prime}(r)
= -\frac{1}{r^{2}} + \e^{-2\lambda} (1-2r\lambda^{\prime})\frac{1}{r^{2}}
\, .
\end{align}
Eq.~(\ref{tov1-1}) can be rewritten in terms of the mass parameter $M(r)$ as
\begin{align}
\frac{2G}{r^{2}}M^{\prime}(r)
=& 8 \pi  G \rho (r)
+ m^{2}_{0}
\left[ 3 - \frac{2 \chi (r)}{r}  - \chi^{\prime} (r) \left( 1 -
\frac{2GM(r)}{r}\right)^{1/2} \right]
\nonumber \\
GM^{\prime}(r)
=& 4 \pi G \rho (r) r^{2}
+ \frac{1}{2} m^{2}_{0} r^{2}
\left[ 3 - \frac{2 \chi (r)}{r}  - \chi^{\prime} (r) \left( 1 - \frac{2GM(r)}{r}\right)^{1/2} \right] \, .
\label{tov1-2}
\end{align}

On the other hand, when we operate the covariant derivative on
Eq.~(\ref{eom}), we find that
\begin{align}
\nabla_{\mu} \left(G^{\mu \nu} + m^{2}_{0} I^{\mu \nu} \right) = \nabla_{\mu} T^{\mu \nu} \, .
\end{align}
In the General Relativity with $I^{\mu \nu}=0$,
$\nabla_{\mu}T^{\mu \nu}=0$ is automatically derived from the Bianchi
identity
$\nabla_{\mu}G^{\mu \nu}=0$.
Therefore, it is reasonable that $T_{\mu \nu}$ is assumed to be separately
conserved,
and the conservation law $\nabla^{\mu} T_{\mu \nu}=0$ gives
\begin{align}
\phi^{\prime}= - \left( P + \rho \right)^{-1} P^{\prime} \, , \quad
\phi = - \int \left( P + \rho \right)^{-1} P^{\prime} dr \, .
\label{phi_rho_P}
\end{align}
Eq.~(\ref{tov2-1}) can be rewritten in terms of the energy-density $\rho$
and the pressure $P$ as
\begin{align}
8 \pi G P(r)
=& - \frac{1}{r^2}
+ \frac{1}{r^2} \left[1 - 2(P+\rho)^{-1}P^{\prime}r \right] \left(1 -
\frac{2GM(r)}{r} \right)
+ m^{2}_{0} \left( 3 - \frac{2 \chi (r)}{r} - \e^{\int \left( P + \rho \right)^{-1} P^{\prime} dr}  \right) \, .
\label{tov2-2}
\end{align}
Furthermore, Eq.(\ref{tov3-1}) is given by
\begin{align}
8 \pi G P(r) =&
\left[ - \left( \left( P + \rho \right)^{-1} P^{\prime} \right)^{\prime}
+  (\left( P + \rho \right)^{-1} P^{\prime})^{2}
- \frac{1}{r}\left( P + \rho \right)^{-1} P^{\prime}  \right] \left( 1 - \frac{2GM(r)}{r} \right)
\nonumber \\
&
- \frac{1}{2} \left[ \left( P + \rho \right)^{-1} P^{\prime} - \frac{1}{r}
\right] \left( 1 - \frac{2GM(r)}{r} \right)^{\prime}
\nonumber \\
&
+ m^{2}_{0} \left( 3 - \frac{\chi (r)}{r}  - \chi^{\prime} (r) \left( 1 -
\frac{2GM(r)}{r} \right)^{1/2}
- \e^{\int \left( P + \rho \right)^{-1} P^{\prime} dr} \right) \, .
\label{tov3-2}
\end{align}

Additionally, the interaction term $I_{\mu \nu}$ has to be separately
conserved,
$\nabla_{\mu} I^{\mu \nu}=0$,
because $\nabla_{\mu}G^{\mu \nu}=0$ and $\nabla_{\mu}T^{\mu \nu}=0$.
Note that we can express the interaction terms
as $I^{\mu \nu} = X^{\mu}_{\lambda}g^{\lambda \nu}$,
where $X^{\mu}_{\lambda}$ is defined as
\begin{align}
X^{\mu}_{\lambda}
=& \mathrm{diag} \left(
3 - \frac{2 \chi (r)}{r} - \chi^{\prime} (r) \e^{-\lambda (r)} ,
3 - \frac{2 \chi (r)}{r} - \e^{-\phi (r)} ,
\right. \nonumber \\
& \left.
3 - \frac{\chi (r)}{r} - \chi^{\prime} (r) \e^{-\lambda (r)} - \e^{-\phi (r)} ,
3 - \frac{\chi (r)}{r} - \chi^{\prime} (r) \e^{-\lambda (r)} - \e^{-\phi(r)}  \right) \, .
\end{align}

Thus, the constraint is written as $\nabla_{\mu}X^{\mu}_{\lambda}=0$,
and a non-trivial relation is given by $\nabla_{\mu}X^{\mu}_{r}=0$ as
\begin{align}
0 =&
\left( \frac{2}{r} + \phi^{\prime}(r) \right)\chi^{\prime}(r) \e^{-\lambda} - \frac{2\chi^{\prime}(r)}{r}
\nonumber \\
=&
\left( \frac{2}{r} - \left( P + \rho \right)^{-1} P^{\prime} \right)
\left( 1 - \frac{2GM(r)}{r} \right)^{1/2}- \frac{2}{r} \, .
\label{bianchi}
\end{align}

Now, we introduce the dimensionless variables defined by
\begin{align}
M \rightarrow mM_{\odot}, \quad r \rightarrow r_{g}r, \quad
\rho \rightarrow \tilde{\rho} M_{\odot}/r^{3}_{g} , \quad P \rightarrow
pM_{\odot}/r^{3}_{g} ,\quad
m_{0} \rightarrow \alpha M_{\odot} \, .
\end{align}
Here $M_{\odot}$ is the solar mass and $r_{g}=GM_{\odot}$.
After the short calculation, Eqs.~(\ref{tov1-2}), (\ref{tov2-2}), and
(\ref{tov3-2}) are rewritten as
\begin{align}
m^{\prime}(r)
=& 4 \pi \tilde{\rho} (r) r^{2}
+ \frac{1}{2} \alpha^{2} \left(r_{g}M_{\odot} \right)^{2} r^{2}
\left[ 3 - \frac{2 \chi (r)}{r}  - \chi^{\prime} (r) \left( 1 - \frac{2m(r)}{r}\right)^{1/2} \right] \, ,
\label{tov1-3}
\end{align}
\begin{align}
8\pi p(r) =&
- \frac{1}{r^2}
+ \frac{1}{r^2} \left[1 - 2(p+\tilde{\rho})^{-1}p^{\prime}r \right]
\left(1 - \frac{2m(r)}{r} \right)
+ \alpha^{2} \left(r_{g}M_{\odot} \right)^{2}
\left( 3 - \frac{2 \chi (r)}{r} - \e^{\int \left( p + \tilde{\rho} \right)^{-1} p^{\prime} dr} \right) \, ,
\label{tov2-3}
\end{align}
\begin{align}
8 \pi p(r) =&
\left[ - \left( \left( p + \tilde{\rho} \right)^{-1} p^{\prime} \right)^{\prime}
+  (\left( p + \tilde{\rho} \right)^{-1} p^{\prime})^{2}
- \frac{1}{r}\left( p + \tilde{\rho} \right)^{-1} p^{\prime}  \right]
\left( 1 - \frac{2m(r)}{r} \right)
\nonumber \\
&
- \frac{1}{2} \left[ \left( p + \tilde{\rho} \right)^{-1} p^{\prime} -
\frac{1}{r} \right] \left( 1 - \frac{2m(r)}{r} \right)^{\prime}
\nonumber \\
&
+ \alpha^{2} \left(r_{g}M_{\odot} \right)^{2}
\left( 3 - \frac{\chi (r)}{r}  - \chi^{\prime} (r) \left( 1 - \frac{2m(r)}{r} \right)^{1/2}
- \e^{\int \left( p + \tilde{\rho} \right)^{-1} p^{\prime} dr} \right) \, .
\label{tov3-3}
\end{align}
And, the constraint (\ref{bianchi}) is rewritten as
\begin{align}
0=&
 \left( - \left( p + \tilde{\rho} \right)^{-1} p^{\prime} + \frac{2}{r}
\right)
\left( 1 - \frac{2m(r)}{r} \right)^{1/2} - \frac{2}{r} \, . \label{bianchi2}
\end{align}
For $m_{0}=0$, Eqs.~(\ref{tov1-3}) and (\ref{tov2-3}) reduce
to ordinary TOV equations consistently,
\begin{align}
m^{\prime}(r)=4\pi \tilde{\rho}(r)r^{2} , \quad
p^{\prime}(r)=\frac{4\pi p(r) r^{3} + m(r)}{r(r-2m(r))}(p(r) +
\tilde{\rho}(r)) \, .
\end{align}

We should note that the constraint equation (\ref{bianchi2}) does not
appear in above equations
because the interaction terms do not appear in the action (\ref{the action}).
We also note that, in Eq.~(\ref{tov2-1}),  $\e^{-\phi}$ appears from the
mass term,
which is a unique property in the massive gravity.
Since $\phi^{\prime}$ can be expressed as a function of $\rho$ and $P$ in
Eq.~(\ref{phi_rho_P}),
the integrations of $\rho$ and $P$ appear in the modified TOV equation.
As we will see in next subsection,
the TOV equation in the massive gravity becomes second-order differential
equation because of the integration.

\subsection{Constraint and field equations}

In the previous subsection, we saw that there appears one constraint equation
coming from the conservation law for the interaction terms.
In this subsection, we substitute the constraint into two fields equations
and complete their further deformation into the form suitable for the
numerical calculation.

At first, the field equations are written as
\begin{align}
m^{\prime}(r)
=& 4 \pi \tilde{\rho} (r) r^{2}
+ \frac{1}{2} \alpha^{2} \left(r_{g}M_{\odot} \right)^{2} r^{2}
\left[ 3 - \frac{2 \chi (r)}{r}  - \chi^{\prime} (r) \left( 1 - \frac{2m(r)}{r}\right)^{1/2} \right] \, , 
\label{eq1}
\end{align}
\begin{align}
8\pi p(r) =&
- \frac{1}{r^2}
+ \frac{1}{r^2} \left(1 - 2qr \right) \left(1 - \frac{2m(r)}{r} \right)
+ \alpha^{2} \left(r_{g}M_{\odot} \right)^{2} \left( 3 - \frac{2 \chi (r)}{r} - \e^{\int q dr} \right) \, ,
\label{eq2}
\end{align}
\begin{align}
8 \pi p(r) =&
\left[ q\left( q  - \frac{1}{r} \right) - q^{\prime} \right] \left( 1 - \frac{2m(r)}{r} \right)
- \frac{1}{2} \left( q - \frac{1}{r} \right) \left( 1 - \frac{2m(r)}{r} \right)^{\prime}
\nonumber \\
&
+ \alpha^{2} \left(r_{g}M_{\odot} \right)^{2}
\left( 3 - \frac{\chi (r)}{r}  - \chi^{\prime} (r) \left( 1 - \frac{2m(r)}{r} \right)^{1/2} - \e^{\int q dr} \right) \, ,
\label{eq4}
\end{align}
and constraint
\begin{align}
0=&
\left( \frac{2}{r} - q \right) \left( 1 - \frac{2m(r)}{r} \right)^{1/2} - \frac{2}{r} \, .
\label{eq3}
\end{align}
Here, we define a new variables $q$ as
\begin{align}
q \equiv&
(p + \tilde{\rho})^{-1}p^{\prime}
\, , \
\tilde{\rho} =
\frac{p^{\prime}}{q} - p \, .
\end{align}

Eq.~(\ref{eq3}) is the constraint by $\nabla_{\mu}I^{\mu \nu}=0$, and can
be rewritten as
\begin{align}
\left( 1 - \frac{2m(r)}{r} \right)^{1/2} = \left( 1 - \frac{1}{2}qr \right)^{-1} \, , 
\end{align}
\begin{align}
m(r) = \frac{1}{2}r - \frac{1}{2} r \left( 1 - \frac{1}{2}q(r)r \right)^{-2} \, ,
\end{align}
\begin{align}
m^{\prime}(r) = 
\frac{1}{2} - \frac{1}{2} \left( 1 - \frac{1}{2}q(r)r \right)^{-2}
- \frac{1}{2}r \left( 1 - \frac{1}{2}q(r)r \right)^{-3} \left( q^{\prime}r + q\right) \, .
\end{align}
Thus, by substituting the Eq.~(\ref{eq3}) into the Eqs.~(\ref{eq1}),
(\ref{eq2}), and (\ref{eq4})
we obtain
\begin{align}
\frac{1}{2} - \frac{1}{2} \left( 1 - \frac{1}{2}qr \right)^{-2}
& - \frac{1}{2}r \left( 1 - \frac{1}{2}qr \right)^{-3} \left( q^{\prime}r + q\right)
\nonumber \\
=& 4 \pi (\frac{p^{\prime}}{q} - p) r^{2}
+ \frac{1}{2} \alpha^{2} \left(r_{g}M_{\odot} \right)^{2} r^{2}
\left[ 3 - \frac{2 \chi (r)}{r}  - \chi^{\prime} (r) \left ( 1 - \frac{1}{2}qr \right )^{-1} \right]
\nonumber \\
8 \pi p(r) =&
8 \pi \frac{p^{\prime}}{q}
- \frac{1}{r^{2}}
+ \frac{1}{r^{2}} \left( 1 - \frac{1}{2}q(r)r \right)^{-2}
+ \frac{1}{r} \left( 1 - \frac{1}{2}q(r)r \right)^{-3} \left( q^{\prime}r
+ q \right) 
\nonumber \\
&
+  \alpha^{2} \left(r_{g}M_{\odot} \right)^{2}
\left[ 3 - \frac{2 \chi (r)}{r}  - \chi^{\prime} (r) \left ( 1 -
\frac{1}{2}qr \right )^{-1} \right] \, ,
\label{eq1-1}
\end{align}
\begin{align}
8\pi p(r) =&
 - \frac{1}{r^2}
+ \frac{1}{r^2} \left(1 - 2qr \right)\left( 1 - \frac{1}{2}qr \right)^{-2}
+ \alpha^{2} \left(r_{g}M_{\odot} \right)^{2}
\left( 3 - \frac{2 \chi (r)}{r} - \e^{\int q dr} \right) \, , 
\label{eq2-1}
\end{align}
\begin{align}
8 \pi p(r) =&
\left[ q\left( q  - \frac{1}{r} \right) - q^{\prime} \right] \left( 1 - \frac{1}{2}qr \right)^{-2}
- \frac{1}{2} \left( q - \frac{1}{r} \right) \left( q^{\prime}r + q \right) \left( 1 - \frac{1}{2}qr \right)^{-3}
\nonumber \\
&
+ \alpha^{2} \left(r_{g}M_{\odot} \right)^{2}
\left[ 3 - \frac{\chi (r)}{r}  - \chi^{\prime} (r) \left( 1 - \frac{1}{2}qr \right)^{-1} - \e^{\int q dr} \right] \, .
\label{eq4-1}
\end{align}

\subsection{Consistency check}

In the last section, we derived the three field equations where the mass
parameter $m(r)$ is eliminated by substituting the constraint.
Here, we have three arbitrary functions; $\chi$, $p$, and $\tilde{\rho}$.
On the other hand, we have three field equations and use one equation of
state later.
So, the system looks as over-constrained,
and we need to check that two equations are identical with each other.

From Eq.~(\ref{eq1-1}) and Eq.~(\ref{eq4-1}),
we can eliminate $\chi^{\prime}$ and obtain
\begin{align}
- 8 \pi \frac{p^{\prime}}{q}
=&
- \frac{1}{r^{2}}
+ \left( \frac{1}{r^{2}} - q \left( q -\frac{1}{r} \right) + q^{\prime}
\right)  \left( 1 - \frac{1}{2}qr \right)^{-2}
\nonumber \\
&
+ \frac{1}{2} \left( q + \frac{1}{r} \right)  \left( q^{\prime}r + q\right) \left( 1 - \frac{1}{2}qr \right)^{-3}
+  \alpha^{2} \left(r_{g}M_{\odot} \right)^{2} \left[  - \frac{\chi (r)}{r} + \e^{\int q dr}  \right] \, .
\end{align}
Furthermore, we use Eq.~(\ref{eq2-1}) and eliminate $\chi$
\begin{align}
8\pi p(r) + 16 \pi \frac{p^{\prime}}{q}
=&
\frac{1}{r^{2}}
- \frac{1}{r^{2}} \left( 2 q^{\prime}r^{2} - 2 q^{2}r^{2} + 4qr + 1 \right)  \left( 1 - \frac{1}{2}qr \right)^{-2}
\nonumber \\
&
- \frac{1}{r}\left( 1 + qr \right)  \left( q^{\prime}r + q \right) \left( 1 - \frac{1}{2}qr \right)^{-3} 
+ 3 \alpha^{2} \left(r_{g}M_{\odot} \right)^{2} \left[ 1 - \e^{\int q dr} \right] \, .
\label{eq11}
\end{align}

On the other hand,
from Eq.~(\ref{eq2-1}), we obtain
\begin{align}
\alpha^{2} \left(r_{g}M_{\odot} \right)^{2} \chi^{\prime}
=& - 4\pi p^{\prime}r - 4\pi p
- \frac{1}{2r^{2}}\left( 1 - \frac{1}{2}qr \right)^{-2} \left( 1- 2qr \right)
\nonumber \\
&
+ \frac{1}{2r}\left( 1 - \frac{1}{2}qr \right)^{-3} \left(q^{\prime}r +q\right) \left( 1- 2qr \right)
- \frac{1}{r}\left( 1 - \frac{1}{2}qr \right)^{-2} \left(q^{\prime}r +q\right)
\nonumber \\
&
+ \frac{1}{2r^{2}}
+ \frac{1}{2} \alpha^{2} \left(r_{g}M_{\odot} \right)^{2} \left[ 3 - (1+qr) \e^{\int q dr}  \right] \, ,
\end{align}
and substitute it into Eq.~(\ref{eq1-1})
\begin{align}
8 \pi p =&
8 \pi \frac{p^{\prime}}{q}
- \frac{1}{r^{2}}
+ \frac{1}{r^{2}} \left( 1 - \frac{1}{2}q(r)r \right)^{-2}
+ \frac{1}{r} \left( 1 - \frac{1}{2}q(r)r \right)^{-3} \left( q^{\prime}r + q \right)
+  3 \alpha^{2} \left(r_{g}M_{\odot} \right)^{2}
\nonumber \\
&
+ 8\pi p
- \frac{1}{r^{2}}\left( 1 - \frac{1}{2}qr \right)^{-2} \left( 1- 2qr \right)
+ \frac{1}{r^{2}}
- \alpha^{2} \left(r_{g}M_{\odot} \right)^{2} \left[ 3 - \e^{\int q dr} \right]
\nonumber \\
&
+ \left( 4\pi p^{\prime}r + 4\pi p
+ \frac{1}{2r^{2}}\left( 1 - \frac{1}{2}qr \right)^{-2} \left( 1- 2qr \right)
\right.
\nonumber \\
&
- \frac{1}{2r}\left( 1 - \frac{1}{2}qr \right)^{-3} \left(q^{\prime}r + q\right) \left( 1- 2qr \right)
+ \frac{1}{r}\left( 1 - \frac{1}{2}qr \right)^{-2} \left(q^{\prime}r + q\right)
\nonumber \\
& \left.
- \frac{1}{2r^{2}}
- \frac{1}{2} \alpha^{2} \left(r_{g}M_{\odot} \right)^{2} \left[ 3 - (1+ qr) \e^{\int q dr}  \right] \right)
\left ( 1 - \frac{1}{2}qr \right )^{-1} \, .
\end{align}

After tedious calculation, we obtain the following equation,
\begin{align}
8\pi p + 16 \pi \frac{p^{\prime}}{q} =&
\frac{1}{r^{2}}
- \frac{1}{r^{2}} \left( 2q^{\prime}r^{2} - 2q^{2}r^{2} + 4qr + 1 \right)
\left( 1 - \frac{1}{2}qr \right)^{-2}
\nonumber \\
&
- \frac{1}{r}\left( 1 - \frac{1}{2}qr \right)^{-3} \left(q^{\prime}r + q\right) \left( 1 + qr \right)
+ 3 \alpha^{2} \left(r_{g}M_{\odot} \right)^{2} \left[ 1 - \e^{\int q dr} \right]
\label{eq12} \, .
\end{align}
Eqs.~(\ref{eq11}) and (\ref{eq12}) are degenerate,
and the number of functions and that of equations are identical with each
other.

Finally, we find that we need to solve the following equation,
\begin{align}
8\pi pq + 16 \pi p^{\prime} =&
 \frac{q}{r^{2}}
- \frac{1}{r^{2}} \left( 2qq^{\prime}r^{2} - 2q^{3}r^{2} + 4q^{2}r + q \right) \left( 1 - \frac{1}{2}qr \right)^{-2}
\nonumber \\
&
- \frac{q}{r}\left( 1 - \frac{1}{2}qr \right)^{-3} \left(q^{\prime}r + q\right) \left( 1 + qr \right)
+ 3 \alpha^{2} \left(r_{g}M_{\odot} \right)^{2} \left[ q - q \e^{\int q dr} \right] \, . 
\label{fieldeq}
\end{align}
For the equation of state which relates the energy density $\rho(r)$ with
the pressure $p(r)$,
Eq.~(\ref{fieldeq}) is expressed as the differential equation with respect
to $p(r)$.
By solving the differential equation, we can find the $r$-dependence of
the pressure $p(r)$.

\section{Numerical analysis for relativistic stars}

\subsection{Quark stars}

In the last section, the fundamental equations and the dynamical variables
have been formulated
to investigate the relativistic star in the dRGT massive gravity.
In this section, we study the quark star and neutron star by numerical
simulations.
Such stars have been studied in the $F(R)$ gravity
\cite{Astashenok:2013vza,
Capozziello:2015yza,Astashenok:2014dja,Capo1,Kostas}.
We compare our results in massive gravity with those in $F(R)$ gravity as
well as in the General Relativity.

The methodology in our numerical simulations is discussed below.
In our study, we solve the third order ordinary differential equation
with three boundary conditions.
We set the first conditions as a value of central density by hand
because we need relations of some parameters for certain central density region.
And, we should the second condition as $p'(r=0) = 0$.
Finally, we set the third condition as that the radius of star becomes identical
with that in the General Relativity for choiced central density.
The radius of the star $r=r_{0}$ is defined by the condition $p(r_{0})=0$,

For solving the equation numerically,
we treat the problem as initial-value problem at the center $r=0$.
We set two of the initial conditions as a value of central density and $p'(r=0) = 0$ as before.
But we should choose the value of $p''(r=0)$ such the last boundary condition is satisfied.
So we optimize the value of $p''(r=0)$ (shooting method).

We should note that the only free parameter in our model is the graviton mass,
and we assume $m_0=\sqrt{\Lambda}$ because we use the dRGT massive gravity.
Also, the graviton mass is constrained by the observation in solar system
and should be small
\cite{Goldhaber:1974wg,Talmadge:1988qz,Will:1997bb,Finn:2001qi,Choudhury:2002pu,Dubovsky:2009xk}.

At first, we study the quark star, using the equation of state called the
MIT bag model \cite{Astashenok:2014dja}.
The equation of state is given by
\begin{align}
p=c(\rho-4B) \, .
\end{align}
Here, $c$ depends on the mass of strange quark $m_{s}$, and $c=0.28$ 
if we choose $m_{s}=250 \mathrm{[MeV]}$.
$B$ is called a bag constant which we fix as $B=60\mathrm{MeV/fm^3}$.
Then, we search $\rho(r)$ which satisfies $p(r)=0$.
Note that the equation of state is linear, thus, we stop the calculation
if $p(r)$ gets negative
and define $r=r_{0}$ as the point between $p(r)>0$ and $p(r)<0$.

\begin{figure}[htbp]
\begin{center}
\includegraphics[width=0.5\hsize]{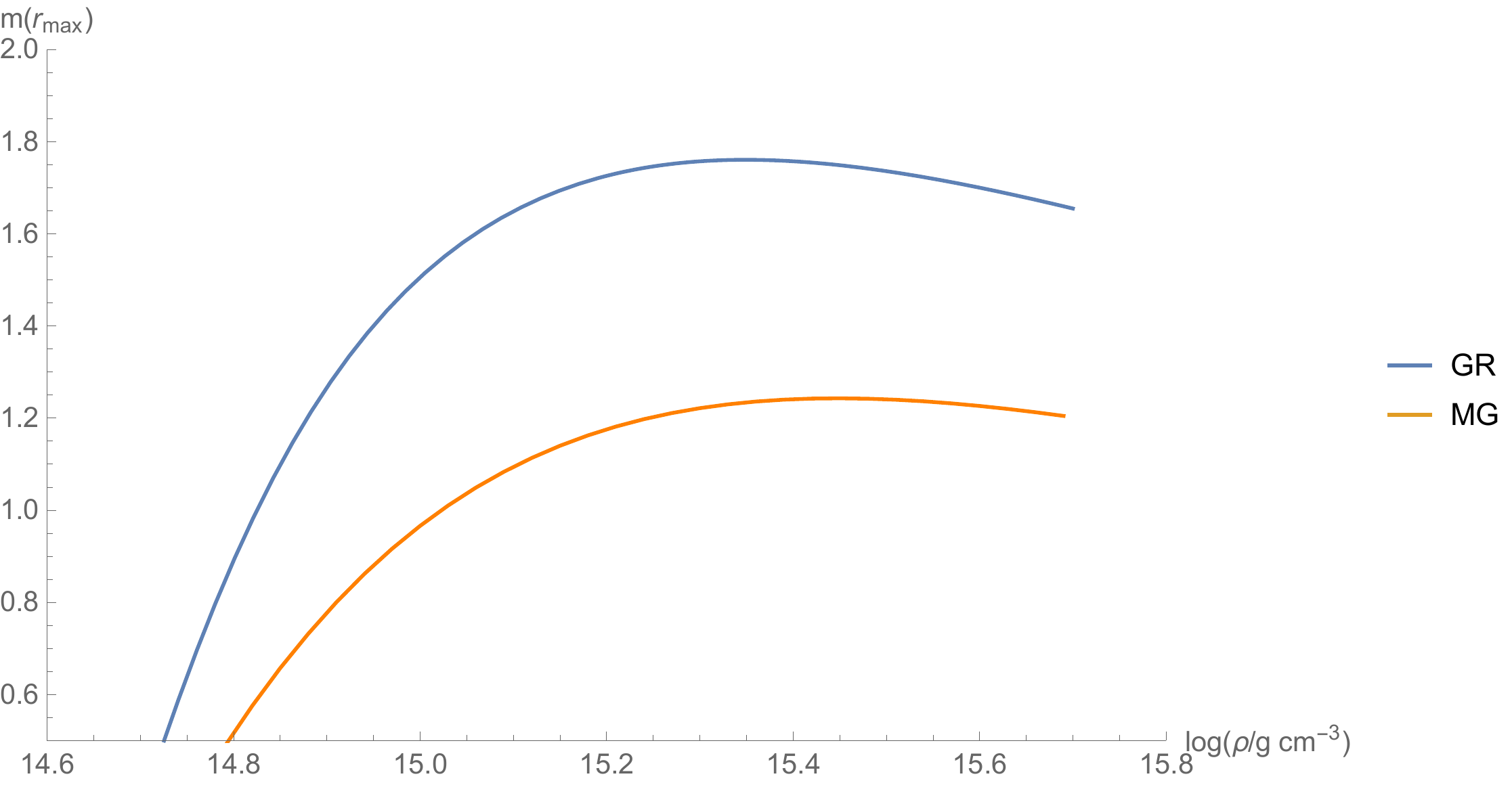}
\caption{Mass-central density relations in the General Relativity
and the massive gravity are shown.}
\label{fig3}
\end{center}
\end{figure}

\begin{figure}[htbp]
\begin{center}
\includegraphics[width=0.5\hsize]{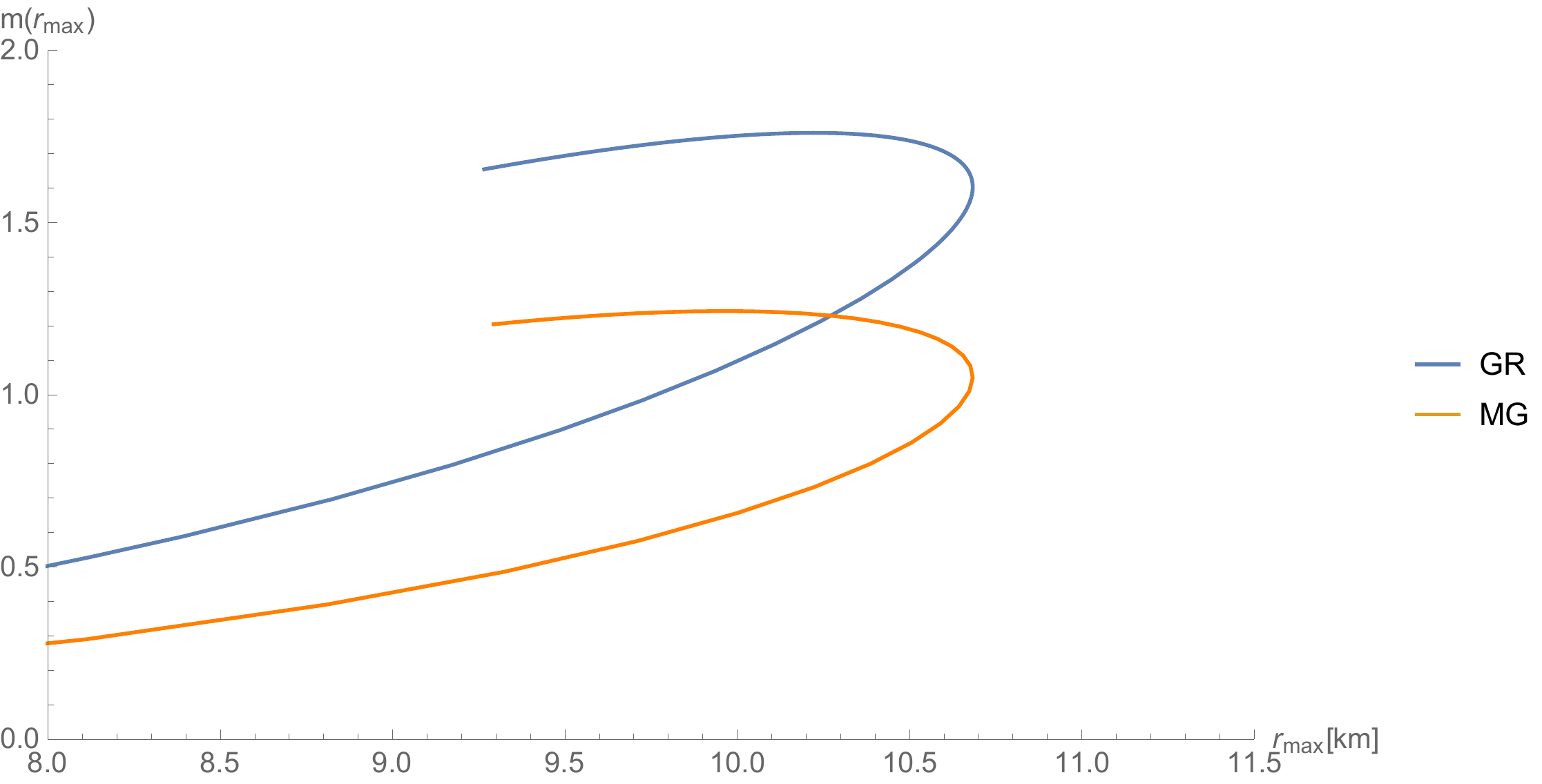}
\caption{Mass-radius relations in the General Relativity and in the massive
gravity are shown.}
\label{fig4}
\end{center}
\end{figure}

We plot $m$-$\tilde{\rho}$, $m$-$r_{\mathrm{max}}$ relations in the
General Relativity (blue line)
and the massive gravity (orange line) (see Figs.~\ref{fig3} and \ref{fig4}).
The total mass of the quark star is smaller than that in the General
Relativity.

\subsection{Neutron stars}

\begin{figure}[htbp]
\begin{center}
\includegraphics[width=0.5\hsize]{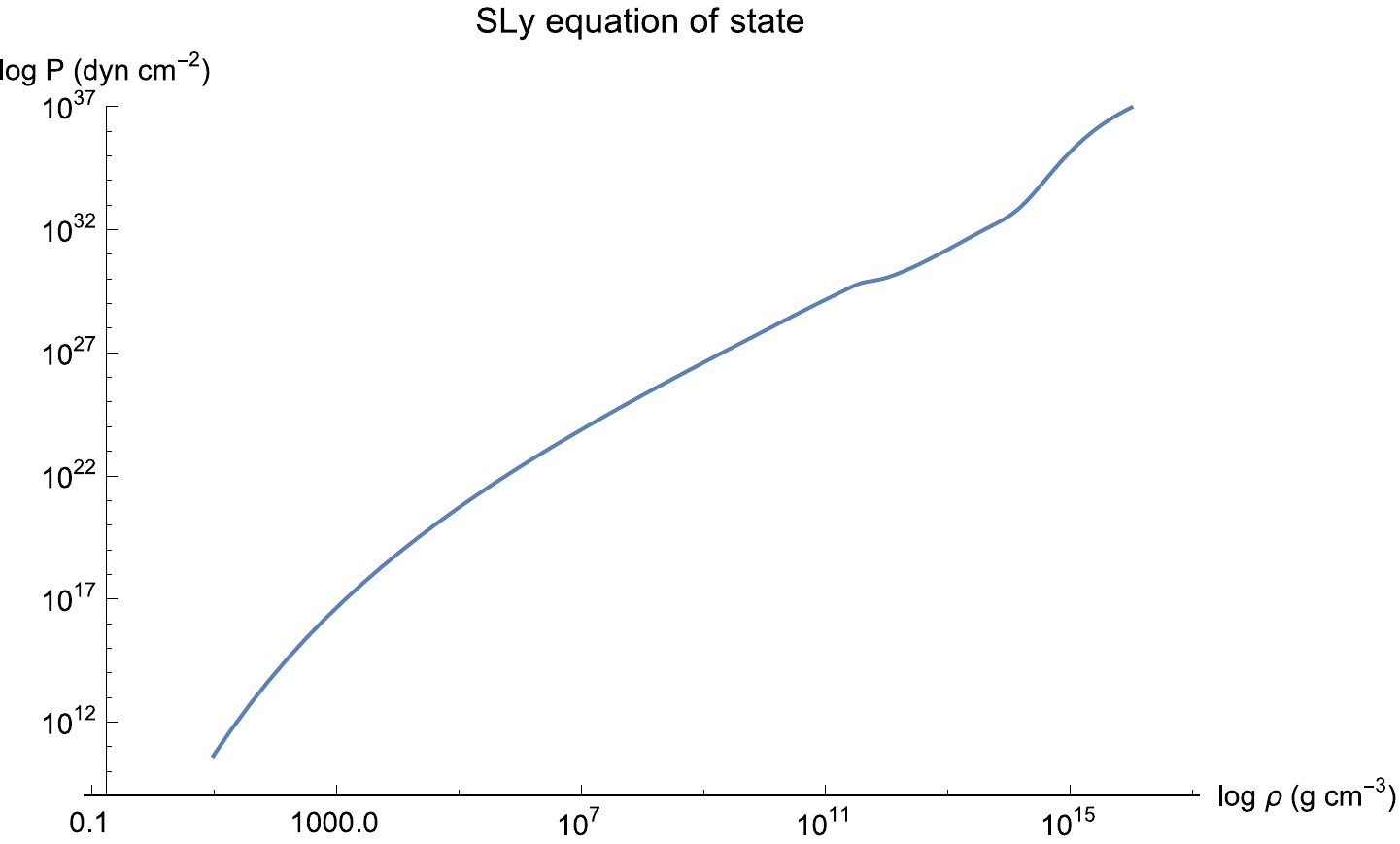}
\caption{SLy equation of state is shown.}
\label{sly}
\end{center}
\end{figure}

Next, we study the neutron stars, using the equation of state called SLy
model
\cite{Haensel:2004nu}.
The equation of state is given by
\begin{align}
\xi =& \log(\rho/\mathrm{g\,cm^{-3}})\, , \quad
\zeta = \log(P/\mathrm{dyn\,cm^{-2}})\, , \quad
f_{0}(x) = \frac{1}{\e^x+1}\, ,
\end{align}
\begin{align}
\zeta =& \frac{6.22 + 6.121\,\xi + 0.006004\,\xi^3}{1 +
0.16345\,\xi}f_0(6.50\,(\xi-11.8440)) \nonumber \\
&\quad+(17.24 + 1.065\,\xi)f_0(6.54\,(11.8421-\xi)) \nonumber \\
&\quad+(-22.003+1.5552\,\xi)f_0(9.3\,(14.19-\xi)) \nonumber \\
&\quad+(23.73 - 1.508\,\xi)f_0(1.79\,(15.13-\xi))\, .
\end{align}
In this case, we utilize a function fitted to numerical points in SLy
model (see Fig.~\ref{sly}),
which is valid if $\rho \geq 10^{5} \mathrm{[g/cm^{3}]}$.
We should note that the equation of state includes logarithmic function,
and the radius is ill-defined if $p(r)<0$.
Therefore, in this case, we define the radius $r=r_{0}$ as
$p(r_{0})<10^{-11}$ and remove ill-defined points.

\begin{figure}[htbp]
\begin{center}
\includegraphics[width=0.5\hsize]{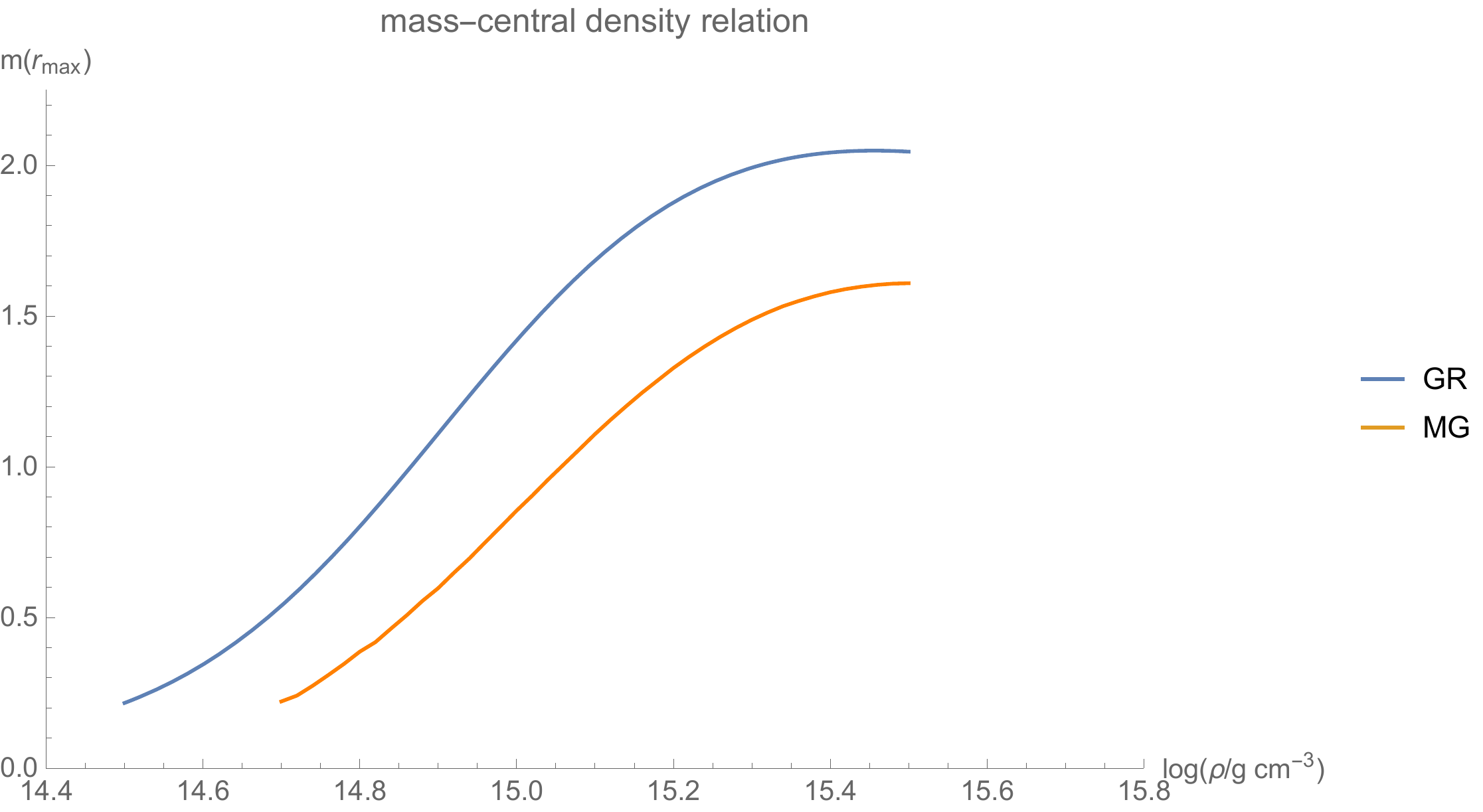}
\caption{Mass-central density relations in the General Relativity and the
massive gravity are shown.}
\label{fig9}
\end{center}
\end{figure}

\begin{figure}[htbp]
\begin{center}
\includegraphics[width=0.5\hsize]{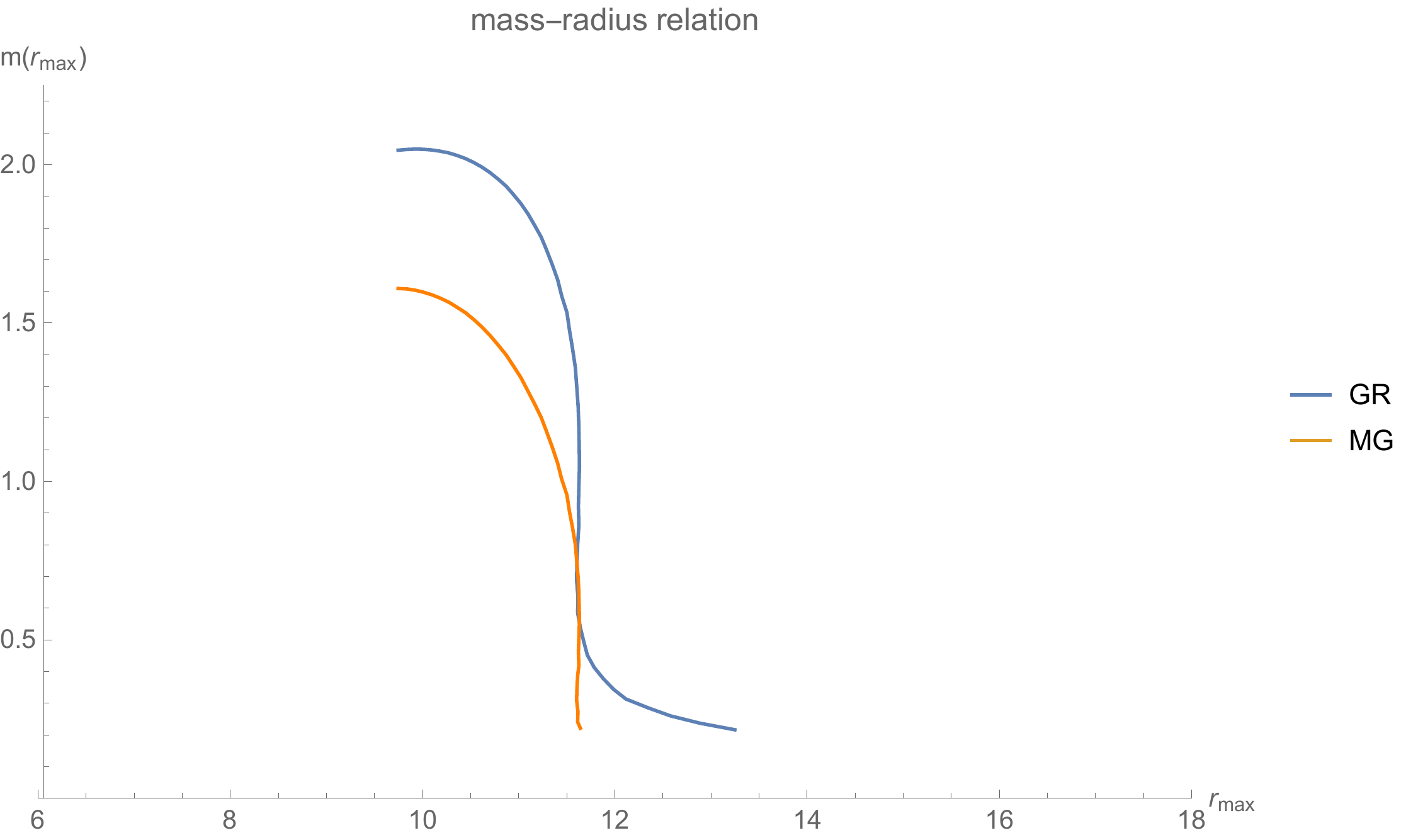}
\caption{Mass-radius relations in the General Relativity and the massive
gravity are shown.}
\label{fig10}
\end{center}
\end{figure}

We plot $m$-$\tilde{\rho}$, $m$-$r_{\mathrm{max}}$ relations (see
Figs.~\ref{fig9} and \ref{fig10}),
where we interpolate lines between plotted points
after we remove the ill-behaved points.
The region of total mass is narrow compared with the case in the General
Relativity,
that is, massive neutron star cannot be realized.

\section{Summary and discussion}

We have investigated the relativistic stars in the dRGT massive gravity
which is non-linear theory of massive graviton.
We assumed the perfect fluid in hydrostatic equilibrium with standard equation of state,
and after we specified the parameters and reference metric
and assumed that graviton mass is comparable with the cosmological constant,
we derived the mass-central density and mass-radius profile for quark
stars and neutron stars numerically.

We have concluded that the TOV equation is corrected by the term
which is proportional to the graviton mass which results from the
potential term of massive graviton in the action,
and one constraint equation appears if we assume the conservation of
energy-momentum tensor.
The correction is very small if we consider the light graviton mass
against massive object.
The mass-radius relation is more constrained than that in the General Relativity,
and we have shown that the maximal mass gets smaller for quark star and neutron star.
Therefore, the compact massive neutron star cannot be explained in this
specific version of dRGT massive gravity,
which is in difference with $F(R)$ gravity
\cite{Capo1,Capozziello:2015yza,Astashenok:2013vza}.
However, the results of our work do not indicate
that the dRGT massive gravity should be excluded by the observation.
Indeed, we have assumed the standard equations of state to study the maximal mass
as well as particular minimal version of massive gravity.
It could be that the massive neutron star could be possible
if we choose the different equations of state or consider more complicated
version of massive gravity with more parameters.

Although the massive neutron star cannot be realized,
we could distinguish the dRGT massive gravity from the General Relativity.
The mass-central density or mass-radius relation for the quark star shows
that the deviation from the General Relativity is very small.
However, for the neutron star, quite important difference appears in the case of small and large radius.
This may be an evidence to quantify the difference from the General Relativity in strong-gravity regime.
This deviation is considered to be derived from the constraint equation (\ref{eq3})
which relates the energy density and pressure inside the neutron star.
On the same time,
the effect of the mass term is very small,
and it is given by the ratio between the graviton mass and neutron star mass $\sim \mathcal{O}(\alpha^{2})$.
The mass term affects the geometry outside the star, 
and causes the accelerated expansion of the Universe at large scales.
Note that the constraint does not depend on the mass of graviton.

On the other hand, the neutron stars in $F(R)$ gravity have been studied
in order to test it in astrophysical phenomenology.
In previous study, the gravitational action is assumed to be $F(R) = R + h(R)$,
where the function $f(R)$ is corresponding to the deviation from the General Relativity.
The mass-central density and mass-radius relation was studied as well as in our work,
and it has been shown that, for specific function of $f(R)$,
the massive neutron star whose mass $M \sim 2M_{\odot}$ is realized.

In this work, we did not studied all models of the dRGT massive gravity
because we fixed the free parameters $\beta_{n}$ and the reference metric
$f_{\mu \nu}$.
Thus, we have two simple ways to generalize our work.
First, one can change the parameters $\alpha_{3}$ and $\alpha_{4}$ in
2-parameter family of the dRGT massive gravity.
It means that the potential of massive graviton is changed and it would
lead to the different mass-radius relation.
Second, we can change the reference metric from the Minkowski to the other ones.
As we mentioned, we have considered the specific class of the reference metric
and formulated the modified TOV equation.
Flat reference metric brings simple but non-trivial modification to the TOV equation.
The reference metric may be chosen in more general form, 
so that it may not be solution of equations of motion in General Relativity.
Regarding the choice of the reference metric, we can also generalize our
study to the massive bigravity theory
which describes the interaction between the massless spin-$2$ field,
corresponding to usual massless graviton,
and massive spin-$2$ field \cite{Hassan:2011zd,Hassan:2011tf,Hassan:2011vm}
(for the $F(R)$ gravity extension of the massive gravity,
see \cite{Nojiri:2012zu,Kluson:2013yaa,Manos}).
In bigravity, the reference metric is dynamical and determined by the equation of motion, thus,
we need not to fix the reference metric by hands.

Finally, we make some remarks about the compact stars in the dRGT massive gravity.
The compact stars usually have a maximal mass, and then, 
the gravitational collapse to the black hole is inevitable.
If the horizon is formed, the interaction terms diverge at the horizon because of $\sqrt{g^{-1}f}$.
This singularity is not removable, therefore, the naked singularity appears.
However, we do not know what happens after the gravitational collapse, precisely.

In the massive gravity, the basic equation forms as 
$G_{\mu \nu} +m^{2}_{0} I_{\mu \nu} = \kappa^{2}T_{\mu \nu}$.
For example, in the case of the Schwarzschild metric, 
the Einstein tensor is equal to zero, $G_{\mu \nu} = 0$.
If the energy-momentum tensor $T_{\mu \nu} = 0$, the interaction terms $I_{\mu \nu}$ should be zero
although these terms include non-trivial effect caused by $\sqrt{g^{-1}f}$
except for the case that $f_{\mu \nu}$ is proportional to the Schwarzschild metric.
For general space-time with horizon, 
if the interaction terms diverge at horizon, $G_{\mu \nu}$ and/or $T_{\mu \nu}$ should diverge
although the strength of gravitational force, curvature, is finite at
horizon, thus, the energy-density and pressure are also finite.
From the view point of analogy to the classical mechanics, the mass term
is a potential of the metric,
and the solution cannot arrive at horizon if potential goes to positive
infinity at horizon.
For the above reasons, the metric with horizon may not be a solution.

The dRGT massive gravity has the cutoff scale $\Lambda_{3}$
where the ghost-mode caused by the higher derivatives is suppressed.
$\Lambda_{3}$ depends on the graviton mass $m_{0}$,
and the cutoff scale is low if the graviton mass is small.
In this work, we have assumed $m_{0} = \sqrt{\Lambda}$, then,
$\Lambda_{3} \sim 1000 \mathrm{[km]}$ which is larger than the scale of
compact stars.
Heavier massive graviton makes the cutoff scale higher,
but it spoils the motivation to explain the late-time acceleration as
modified gravity.
If we accept the cosmological constant and heavy graviton mass,
we can explain the relativistic stars and
our numerical results do not change drastically
because the correction is proportional to the ratio between graviton mass
and massive star and it is still small.

\section*{Acknowledgments}

This research is supported in part by
MEXT KAKENHI Grant-in-Aid for Scientific Research on Innovative Areas
``Cosmic
Acceleration'' (No. 15H05890) and the JSPS Grant-in-Aid for Scientific
Research (C) \# 23540296(S.N.), by the Grant-in-Aid for JSPS Fellows \#
15J06973(T.K.) and by MINECO (Spain) project FIS2013-44881 and I-LINK 1019
and
JSPS fellowship ID No.:S15127(S.D.O.).

\end{document}